\documentclass[%
reprint,
aip,
bibnotes,
 amsmath,amssymb,
floatfix,
]{revtex4-2}

\usepackage[english]{babel}
\usepackage[utf8]{inputenc}
\usepackage[colorinlistoftodos, color=green!40, prependcaption]{todonotes}
\usepackage[pdftex, pdftitle={Article}, pdfauthor={Author}]{hyperref} 
\usepackage{siunitx}
\usepackage{physics}
\usepackage{comment}
\usepackage{xcolor}
\begin{document}
\title{Quantifying the Sensitivity and Unclonability of Optical Physical Unclonable Functions}

\author{Giuseppe Emanuele Lio}
    \email[Correspondence email address: ]{lio@lens.unifi.it}
    \affiliation{Nation Institute of Optics National Research Council (CNR-INO), European Laboratory for Non-Linear Spectroscopy (LENS) and Physics Dept., University of Florence, Via Nello Carrara 1, 50019, Sesto Fiorentino, Florence, Italy}
    
\author{Sara Nocentini}
    \affiliation{Istituto Nazionale di Ricerca Metrologica (INRiM), Strada delle Cacce 91, Turin, 10135 Italy and European Laboratory for Non-Linear Spectroscopy (LENS), University of Florence, Via Nello Carrara 1, 50019, Sesto Fiorentino, Florence, Italy}
    
\author{Lorenzo Pattelli}
    \affiliation{Istituto Nazionale di Ricerca Metrologica (INRiM), Strada delle Cacce 91, Turin, 10135 Italy and European Laboratory for Non-Linear Spectroscopy (LENS), University of Florence, Via Nello Carrara 1, 50019, Sesto Fiorentino, Florence, Italy}  
    
\author{Eleonora Cara}
     \affiliation{Istituto Nazionale di Ricerca Metrologica (INRiM), Strada delle Cacce 91, Turin, 10135 Italy}

\author{Diederik Sybolt Wiersma}
    \affiliation{Istituto Nazionale di Ricerca Metrologica (INRiM), Strada delle Cacce 91, Turin, 10135 Italy. European Laboratory for Non-Linear Spectroscopy (LENS) and Physics Dept., University of Florence, Via Nello Carrara 1, 50019, Sesto Fiorentino, Florence, Italy}    
    
\author{Ulrich R\"{u}hrmair}
    \email[Correspondence email address: ]{ruehrmair@ilo.de}
    \affiliation{Physics Dept. LMU M\"{u}nchen, Schellingstraße 4/III D-80799 M\"{u}nchen, Germany. Electrical and Computer Engineering (ECE) Dept., University of Connecticut, Storrs, CT, USA}    
    
\author{Francesco Riboli}
    \email[Correspondence email address: ]{riboli@lens.unifi.it}
    \affiliation{Nation Institute of Optics National Research Council (CNR-INO), and European Laboratory for Non-Linear Spectroscopy (LENS), University of Florence, Via Nello Carrara 1, 50019, Sesto Fiorentino, Florence, Italy}    
\date{\today} 

\begin{abstract}
Due to their unmatched entropy, complexity, and security level, optical Physical Unclonable Functions (PUFs) currently receive a lot of interest in the literature. Despite the large body of existing works, however, one of their core features has never been quantified in detail, namely their physical unclonability. This paper tackles this fundamental and yet largely unaddressed issue. In simulations and/or experiments, the sensitivity of diffraction-based optical responses is investigated with respect to various small alterations such as variation in the position, size, and number of the scatterers, as well as perturbations in the spatial alignment between the physical unclonable function (PUF) and the measurement apparatus. Our analysis focuses on 2D optical PUFs because of their relevance in integrated applications and the need to reply to security concerns that can be raised when the physical structure of the geometry is accessible. Among the results of this study, the sensitivity analysis shows that a positional perturbation of scatterers on the order of \SI{30}{\nano\meter}, i.e., far below the wavelength of the probing laser light of \SI{632}{\nano\meter} wavelength, is sufficient to invalidate the PUF response and thus detect a forgery attempt. These results support and quantify the high adversarial efforts required to clone optical PUFs, even for 2D layouts.
\end{abstract}


\maketitle

\section{Introduction}
According to recent estimates \cite{cisco2020cisco}, the looming internet of things and worldwide information exchange by the year 2018-2023 will produce a global data stream of around tens zettabytes per annum.
This requires secure and reliable authentication methods in order to protect private information and to safeguard access to personal devices and services.
The currently widespread techniques to this end rely on the permanent storage of digital secret keys in electronic devices, for example in smartphones, car keys, bank cards, passports, or computers.
Unfortunately, the last decades have seen an explosion of attacks that can extract such keys unnoticedly, including sophisticated malware and physical methods \cite{anderson2020security,kocher2019spectre,lipp2018meltdown}.
This obviously calls for new authentication approaches with improved security features. 

The use of non-digital primitives such as Physical Unclonable Functions (PUFs) constitutes a promising new avenue in this context \cite{pappu2002physical, gassend2002silicon, ruhrmair2014pufs, herder2014physical}.
PUFs are randomly structured physical systems which exhibit a complex input-output or, in PUF parlance, ``challenge-response'' behavior that is unique to each PUF.
Their uncontrollable individual disorder on small length scales makes them practically unclonable, even for their original manufacturer.
Due to their physical nature, randomness, and unclonability, PUFs can disable various popular attack vectors compared to classical, permanently stored keys: For example, their physical nature obviously prevents that PUFs are stolen remotely over a purely digital data connection by attackers \cite{anderson2020security, ruhrmair2013power}.
As another example, PUFs allow the short-term derivation of individual secret key material in devices, avoiding the it long-term and attack-prone presence of secrets in digital memory. This usually complicates key extraction \cite{anderson2020security} and side channel attacks \cite{ruhrmair2013power}.
Finally, some special, advanced subclass of PUFs can remain practically unclonable and thus secure even if all their internal disorder and structure is known, simply due to the current limits of nanofabrication \cite{pappu2002physical,ruhrmair2022secret}.
This makes these special PUFs innately immune against any key-extracting and even any secret-extracting attacks, an seminal property sometimes referred to as ``secret-freeness'' \cite{ruhrmair2022secret}. 

Within the ample research landscape of magnetic \cite{das2015mram}, silicon \cite{gassend2002silicon, suh2007physical, nguyen2018interpose, lugli2013physical, csaba2010application} or radio-wave based \cite{dejean2007rf} PUFs, optical and photonic versions have played a pivotal role since their first proposal in 2002 \cite{pappu2002physical}. 
A generic optical physical unclonable function consists typically of a scattering material, which generates a complex light diffraction pattern when illuminated with coherent light, providing a particularly sensitive and convenient probing mechanism for such systems \cite{pappu2002physical, vskoric2005robust, tuyls2007security, horstmeyer2013physical}. Laser light can resolve their unique structures with sub-wavelength sensitivity, leading to strong security levels and high resilience against cloning \cite{pappu2002physical, vskoric2005robust, tuyls2007security, horstmeyer2013physical,HOR15}.
By using passive or active materials of different nature (integrated light sources in dielectric or metallic systems), and by playing with the systems' dimensionality (2D or 3D), a large variety of optical PUFs can be conceived, as research in the last decades has demonstrated. This includes PUFs based on organic nano-emitters\cite{feng2019random, kayaci2021organic}, chip-scale laser\cite{kim2021massively}, thin random scattering layers of plasmonic nano-particles\cite{emanuele2020opto, berk2021theory, berk2021tracking, ferraro2022low}, random silver nano-structures\cite{anderson2017initial, caligiuri2021hybrid}, or even PUFs architecture compatible with microfabrication technologies for photonic integrated circuits (PIC) \cite{tarik2020robust, grubel2017silicon, knechtel2019toward}. In the end, any material with random structure, defects or scattering elements, including regular paper\cite{buchanan2005fingerprinting}, will generate complex speckle patterns when illuminated by a coherent source, making optical PUFs a highly efficient, inexpensive and robust platform for secure authentication\cite{pappu2002physical, li2017simulation, anderson2017initial, herder2014physical, maes2010physically}. 
In recent years, 2D optical PUFs have attracted particular attention due to their high stability, industry-compatible fabrication processes, and straightforward integration with existing telecommunication technologies \cite{tarik2020robust, grubel2017silicon, knechtel2019toward}.
At the same time, however, these 2D structures inevitably exhibit a lower complexity and entropy than comparable 3D systems.
They also can be directly inspected by electron microscopy or other diagnostic techniques, and are therefore easier -- at least in principle -- to replicate or ``clone'', both experimentally and numerically in simulations. For these reasons, in view of their future widespread adoption, an accurate estimate of their cryptographic security is fundamental and in this work, we quantitatively evaluate their resilience to cloning attacks and sensitivity to measurement perturbations.
In more detail, the experimental and numerical analysis carried out in this work quantifies the sensitivity and the unclonability of a prototypical 2D optical PUF by comparing the keys generated by different clones of the same primitive, or studying the keys variation to small readout alterations.
Perturbations considered include the imperfect cloning of the PUF layout (e.g., due to slightly incorrect number, position or alignment of the scattering elements), as well as errors during the illumination or readout process. All results are interpreted under a unified framework, which allows us to cast a direct connection between the experimental device and its simulated counterpart.
Moreover, the results obtained from the analysis of 2D PUFs are also relevant to more complex 3D architectures, as they can be considered as a lower bound to evaluate unclonability, stability, and other properties in 3D PUFs. Please recall in this context that with current computational methods, exact simulations of 3D optical PUFs are extremely demanding and time consuming compared to the 2D case.
\section{Background and \\ Methodology}
\begin{figure*}[!ht]
\centering
\includegraphics[width=\linewidth]{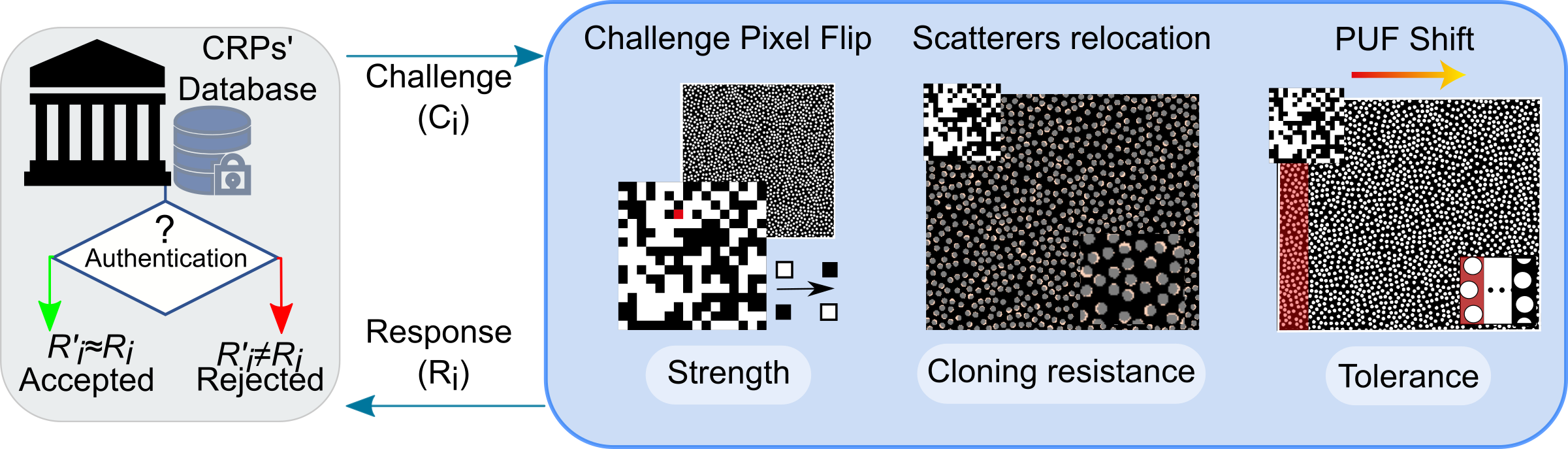}
\caption{a) Schematic representation of the authentication flowchart based on optical PUFs. Potential attack scenarios or issues with this scheme include: i) the challenges $C_i$ are affected by noise (single or multiple macro-pixel flip). ii) the optical-PUF scatterers are misplaced (small structural perturbations or cloning imperfections). iii) the optical-PUF is misaligned with respect to the illumination/readout system.}
\label{fig1}
\end{figure*}
\subsection{Use Case:  Remote Identification}

The typical use case for so-called optical Strong PUFs \cite{mcgrath2019puf} consists of a remote identification protocol, in which a PUF (or a user holding it) identifies remotely via a digital communication channel to a central authority. The protocol employs a large number of input-output pairs or challenge-response pairs (CRPs) of the PUF as a unique identifier or ``fingerprint'', and is sketched in Figure \ref{fig1}. 
As a preparatory initial step, we assume that the central authority has measured a sufficiently large database of CRPs for the PUF.
We also assume the authority has determined some error-tolerating threshold value to ensure a successful identification of the PUF in variable and error-prone everyday conditions.
The threshold must be set is such a way to allow discriminating between the responses of original PUFs and perturbed responses generated by possible read-out errors (Challenge Pixel Flip), non perfect clones of the original PUF (scatterers relocation), or slight misalignment between the original PUF and the read out system (PUF misalignment), see Figure \ref{fig1}.
Once these setup steps have been accomplished, the PUF is handed over from the authority to the user.
During the authentication phase or identification protocol, the authority sends a series of randomly chosen challenges from its database to the PUF/user, and awaits the correct responses in return.
Once they arrive, the incoming responses are compared to the responses measured earlier in the preparatory phase by the authority.
If they match within the predefined error threshold (see above), then the identity of the PUF/user is confirmed.
We stress that each CRP can be used only once in the above protocol.
This means that the pre-established CRP list shrinks over time, and must be planned large enough in the setup phase for the entire application lifetime.
\subsection{Experimental Setup}
\begin{figure}[!ht]
\centering
\includegraphics[width=\columnwidth]{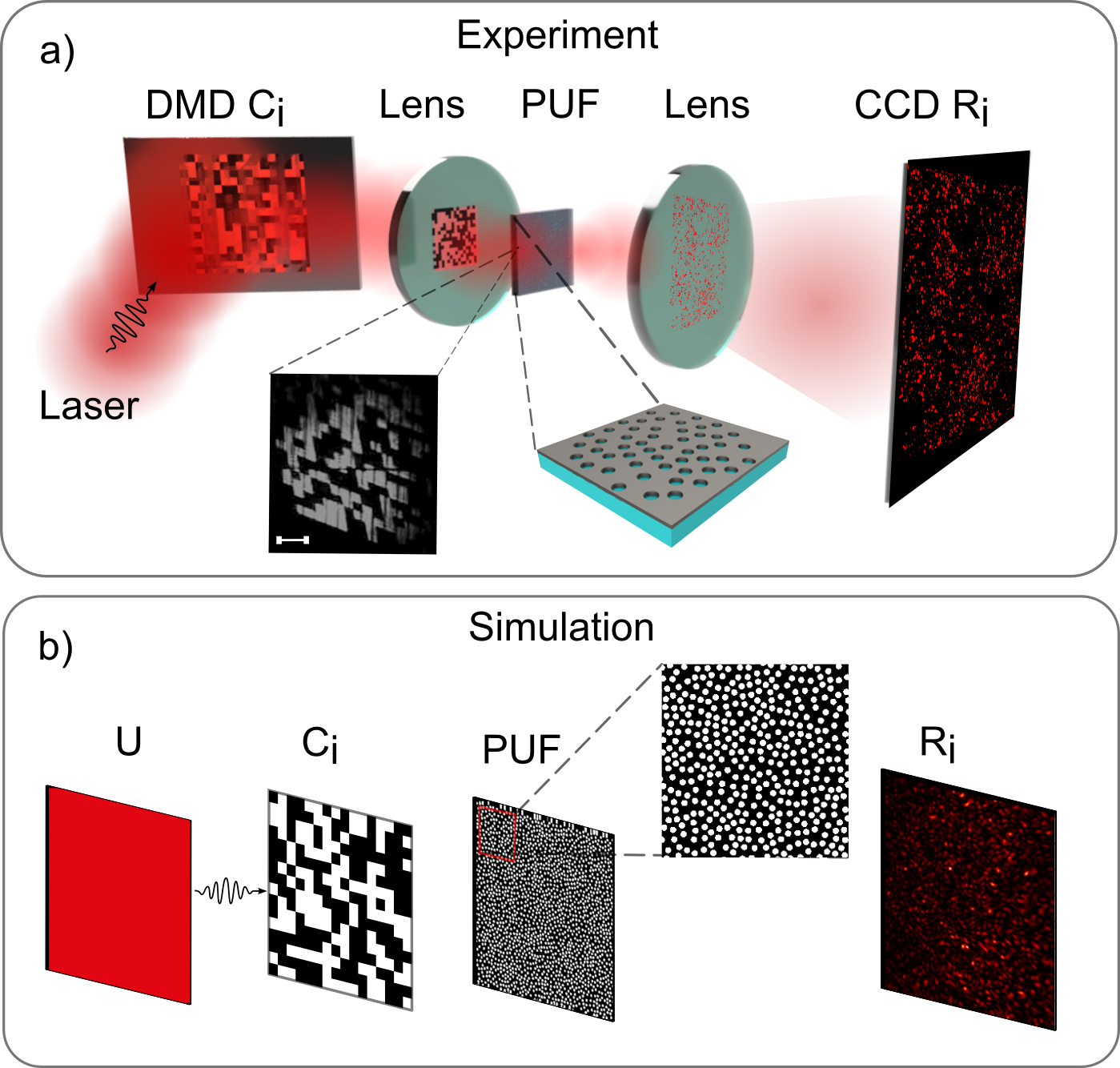}
\caption{a) Sketch of the experimental optical setup used to characterize the proposed optical PUFs. The inset on the left shows the challenges overlapped to the Thorlabs ruler (R1L3S2P) to measure its size (scale bar \SI{100}{\micro\meter}), the inset on the right shows a sketch of the sample. b) Scheme of the numerical workflow used to generate and collect synthetic CRPs.}
\label{fig2}
\end{figure}

To experimentally address these cases, we built an optical setup able to generate the CRPs and authenticate each entity, following the scheme sketched in Figure \ref{fig2}a.
It comprises a He-Ne laser ($\lambda= \SI{633}{\nano\meter}$), a digital micro-mirror device (DMD) for the challenge manipulation, illumination and collection optics, and a CCD camera to record the responses.
Additional details about the experimental apparatus are reported in the Experimental Section and in Figure \ref{FigS1} of the Supporting Information (SI).
The 2D optical-PUF consists of a perforated metallic membrane obtained starting from a disordered 2D arrangement of polystyrene nano-spheres (see insets of Figures \ref{fig2}a and \ref{fig2}b), additional details are available in the Experimental section and in Figure \ref{FigS3} of the SI. Because of the fabrication method based on an uncontrolled self-assembly process, even the manufacturer cannot replicate the same PUF design twice. 
\subsection{Numerical Simulations} 
In parallel, the experimental configuration is replicated numerically using an implementation of the Rayleigh-Sommerfeld (RS) method \cite{torcal2017single, sanchez2018far, sanchez2020optimization, lio2021leveraging}, which we have integrated in a routine to generate the sample masks, the challenges, and a speckle registration algorithm.
The numerical process consists of four main steps, as depicted in Figure \ref{fig2}b.
Numerically, the disordered arrangement of holes in the diffraction mask (i.e., the PUF) is generated by packing non-overlapping circles using either a random sequential adsorption (RSA) or a Lubachevsky-Stillinger (LS) approach \cite{skoge2006packing, riboli2017tailoring}.
These methods are used to pack discs up to a target the packing fraction ($f_\text{p}$) defined as the ratio between the area occupied by the disks and the total sample area.
Then, a plane wave (U) is projected over a pixelated-mask to generate the challenge pattern ($C_i$) which is imaged onto the PUF.
Following the experimental configuration, the scattered intensity is finally recorded in the far field at an off-axis position on a $XY$ plane at a distance $z$ from the PUF.
\subsection{Entropy Estimation for Optical PUFs}
\begin{figure}[!ht]
\centering
\includegraphics[width=\columnwidth]{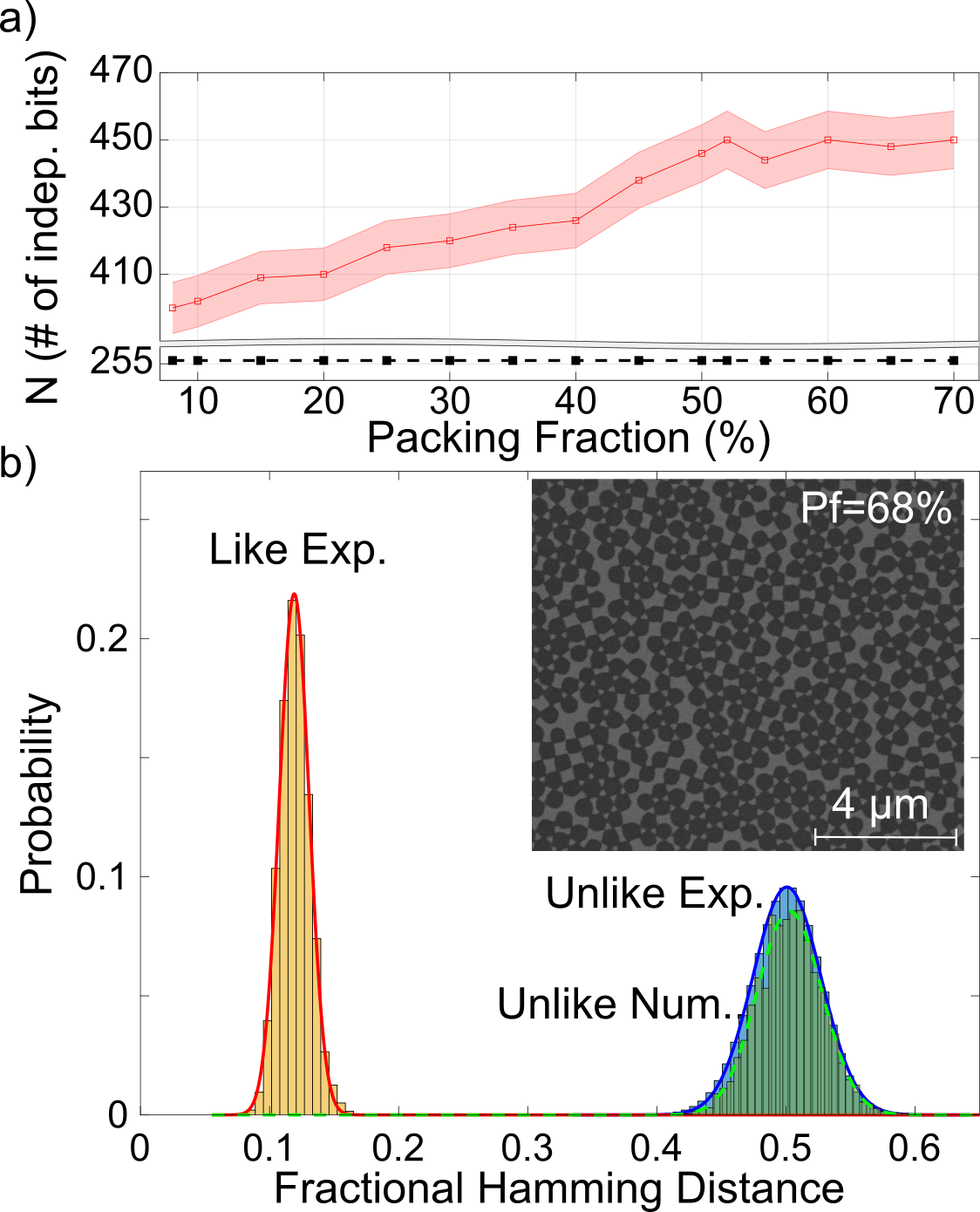}
\caption{a) Number of independent bits calculated as function of $f_\text{p}$ in synthetic optical-PUF samples. b) The blue histogram and solid line correspond to the experimental ``unlike'' and its fit, the experimental ``like'' are displayed with the yellow histogram and the red solid line for the fit, while the numerical ``unlike'' are indicated by the green histogram and its fit by the green dashed line. The inset in b) shows a SEM image of the optical PUF, exhibiting a densely packed arrangement of holes etched into a titanium membrane.}
\label{fig3}
\end{figure}

To quantify the randomness (entropy or information content), and the stability of the PUF responses against environmental variations, we follow the typical approach used for their application \cite{pappu2002physical, daugman2003importance}.
The first step is to generate the binary keys ($K_1, \dots, K_i$ from the responses $R_1, \dots, R_i$ related to the challenges $C_1, \dots, C_i$) by hashing and reshaping each raw speckle image into a 1D array.
This can be performed using standard image transformation and binarization algorithms.
Here, the wavelength of a wavelet-based Gabor filter is tuned to extract the features of the speckle images while ensuring the repeatability of the responses under the same challenge interrogation.
The pairwise distance between each binary keys $K_1, \dots, K_i$ is then measured with the Hamming distance metric \cite{daugman2003importance}.
Distances between keys generated by different challenges are called ``unlike'' distances (and are related to the entropy of the key), while those generated by same challenges are called ``like'' distances (and are related to the stability of the PUF).
The entropy of the keys is then evaluated by assuming that each Fractional Hamming Distance (FHD) resulting from the bit-wise comparison of two different keys can be represented as a Bernoulli trial, albeit with correlations between successive PUF responses.
For large $N$ values, the expected binomial distribution is well approximated by a Gaussian, which makes it possible to estimate the number of independent bit of the keys, i.e., $N=\expval{p} \cdot (1-\expval{p})/\sigma^2$, that is associated with the PUF entropy/information content.
Therefore, as a first step, we have generated several synthetic PUF configurations to study how $N$ depends on the $f_\text{p}$ for a fixed sub-wavelength hole radius $r_0 \approx \SI{200}{\nano\meter}$. Numerical simulations show that dense perforated masks with a $f_\text{p}$ between \SI{50}{\percent} and \SI{70}{\percent} generates keys with the highest entropic content, see Figure \ref{fig3}a. In addition, we note that the entropy of the key is larger than that of the challenge (composed by $M \times M$ macropixels), which is $N_C$ = $\log_2(2)^M$ (see Figure \ref{fig3}a), demonstrating that the interaction between light and the optical-PUF effectively increases the information content encoded in the Challenge. The choice of the size of the macropixels is based on the analysis reported in the Challenges pixel size section and in Figure \ref{FigS2} of the SI.
The optical characterization of the FHD for the experimental PUF sample is reported in Figure \ref{fig3}b, returning a value of $N = 448$ bits, in excellent agreement with the numerical prediction.
The ``like'' FHD distribution, which is collected after a time lag of 30 minutes in order to test the stability and define the acceptance threshold, returns a mean value $\expval{p}$ of $0.122$ and a standard deviation $\sigma = 0.038$ (yellow histogram/red curve in Figure \ref{fig3}b). Since the like and unlike distributions are well separated, the authentication acceptance threshold can be set at around \num{0.2} to safely reject false positives \cite{kim2022revisiting}.
Due to the deterministic and noiseless nature of numerical calculations, the ``like'' FHD distribution is not reported here as it would appear as a delta-distribution centered around zero. Based on the good agreement between the FHD histograms and their Gaussian models, in the following we will plot FHD distributions showing only their fitting curves for better clarity.
\section{Results and Discussion}
\subsection{Sensitivity to Challenge Pixel Flips}
\begin{figure}[!h]
\centering
\includegraphics[width=\columnwidth]{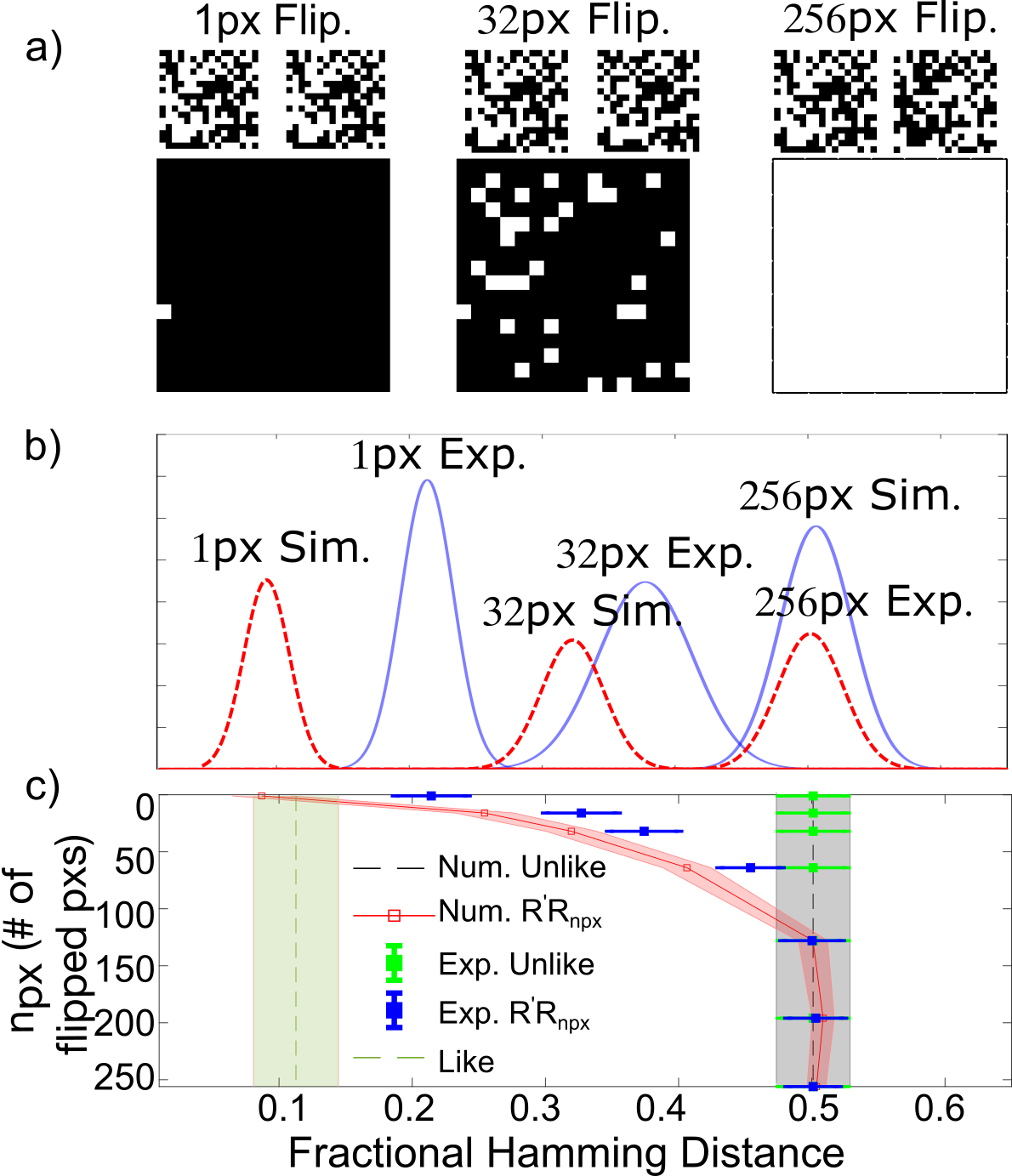}
\caption{a) Illustrative examples of original and perturbed challenges for different numbers of flipped macro-pixels. The large panel shows the macro-pixel that has been flipped. b) Numerical (red) and experimental (blue) ``inter-device'' FHD distributions for 1, 32, and 256 pixels. c) Trend for numerical and experimental ``unlike'' (black ribbon line and green dots), $R'R_\text{npx}$ comparison (red ribbon line and blue scatterers) and the experimental ``like'' (light green ribbon). The error bars are calculated as the standard deviation of each FHD distribution.}
\label{fig4}
\end{figure}
This analysis has been carried out experimentally and verified by numerical calculation.
As a first test, we study the sensitivity of the optical PUF response to perturbations of the challenge (random macro-pixel flips). We generate a set of \num{2000} challenges, and then perturbed versions of these challenges with one or more random macro-pixels flipped to its opposite value. An ideal PUF is expected to provide a completely independent response as soon as the challenge is modified, meaning that this test can be used to evaluate the optical PUF sensitivity. Figure \ref{fig4}a shows some illustrative examples of $16 \times 16$ challenges and their perturbed version, with the flipped pixels drawn in white in the large panels. Figure \ref{fig4}b reports the fit retrieved by the FHD analysis for the experimental and numerical studies. The fits are evaluated on the FHD distributions obtained comparing the responses from the system probed with the original challenges $R'$ and the responses obtained for the perturbed challenges $R_\text{npx}$. Figure \ref{fig4}c summarizes the comparison for different degrees of perturbation. Black and green ribbons are shown for reference, representing the experimental ``unlike'' and ``like'' FHD distributions, respectively. The numerical and experimental  comparison (``inter-device'') FHD distributions of $R'R_\text{npx}$ present a similar trend (red ribbon line and blue dots, respectively). Despite the bi-dimensional nature of the sample, the optical PUF shows a good sensitivity to a single pixel variation of the challenge as the FHD distribution of the responses related to challenges with a single pixel flip are not overlapped with their respective ``like'' distributions.
\subsection{Sensitivity to Scatterers Relocation}
\begin{figure}[!ht]
\includegraphics[width=\columnwidth]{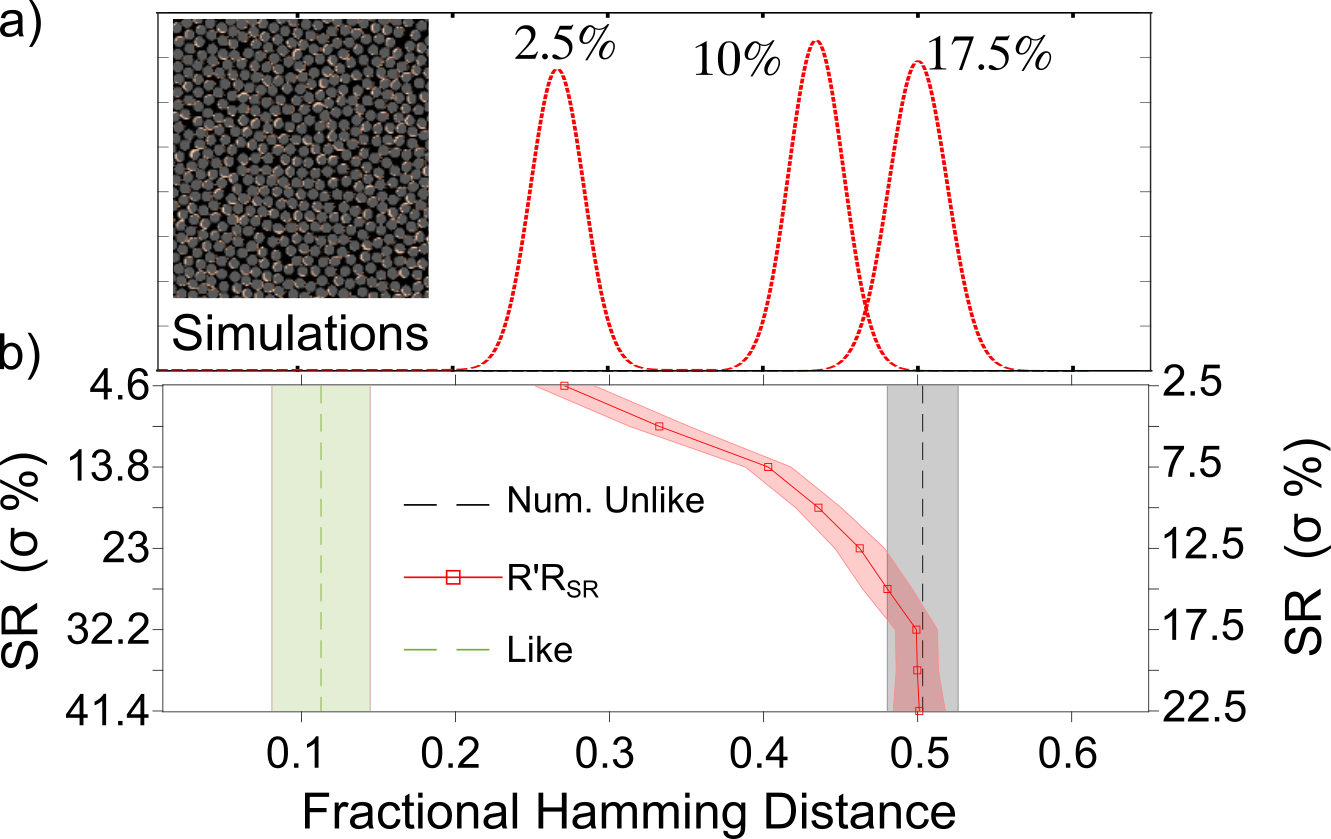}
\caption{a) FHD distributions obtained by perturbing the scatterer positions for $\sigma$ equal to \SI{2.5}{\percent}, \SI{10}{\percent} and \SI{17.5}{\percent}, the inset shows a schematic view of misplaced scatterers. b) FHD trend for the $R'R_\text{shake\%}$ comparison (red ribbon). The experimental ``unlike'' (black ribbon line) and ``like'' distributions (light green ribbon) are also shown for reference.. 
c) Representative sketch about randomly removed and removed and added scatterers in new positions. d) The comparison $R'R_\text{sc\%}$ for different percentages of RR and RA scatterers modifications (red solid and blue dashed lines respectively. e) FHD trend for numerical "unlikes" (black (RR) and green (RA) ribbons), compared FHD distributions $R'R_\text{sc\%}$ (red (RR) and blue ribbons (RA)), and ``like'' distributions (light green ribbon) represents the reference.}
\label{fig5}
\end{figure}  
The second test that can be addressed numerically, concerns the robustness of the PUF to cloning attempts. In this scenario, an attacker trying to clone the physical device aims at replicating the shape, size and position of all scattering elements. To simulate different degrees of cloning imperfections, we perturb the PUF by adding some Gaussian noise ($\sigma \approx n_R/r_0$, where $n_R$ is the relocation) on the position of each scatterer. The results are summarized in Figure \ref{fig5}a, showing that cloning the holes with a precision \SI{<5}{\nano\meter} (below the typical resolution of e-beam lithography fabrications) is already sufficient to relocate the $R'R_\text{SR\%}$ FHD distribution at $\expval{p}=0.25$, well above the experimental acceptance region. With a precision of about \SI{30}{\nano\meter}, the resulting PUF gives rise to an effectively independent set of responses ($\expval{p}=0.5$), which is remarkable considering that the hole radius $r_0$ is $\approx$ \SI{200}{\nano\meter} and the probing wavelength is $\lambda=$ \SI{633}{\nano\meter}. A graph that summarize the behavior for all applied positioning uncertainties is shown in Figure \ref{fig5}b.
Further cloning imperfections are represented by two case studies mimicking an increasing modification of the involved scatterers, for this purpose they are randomly removed (RR) or randomly removed and then added again (RA) in new positions, as sketched in Figure \ref{fig5}c. These numerical studies highlight how the ``inter-device'' (comparison) FHD distributions become independent after a RR\ or a RA\ of \SI{<20}{\percent} ($\expval{p} \approx 0.30$) well above the experimental acceptance region, as shown in the fits and the summary FHD trend in Figure \ref{fig5}d,e respectively. 
\subsection{Sensitivity to Misalignment within Measurement Setup}
\begin{figure}[!ht]
\centering
\includegraphics[width=\columnwidth]{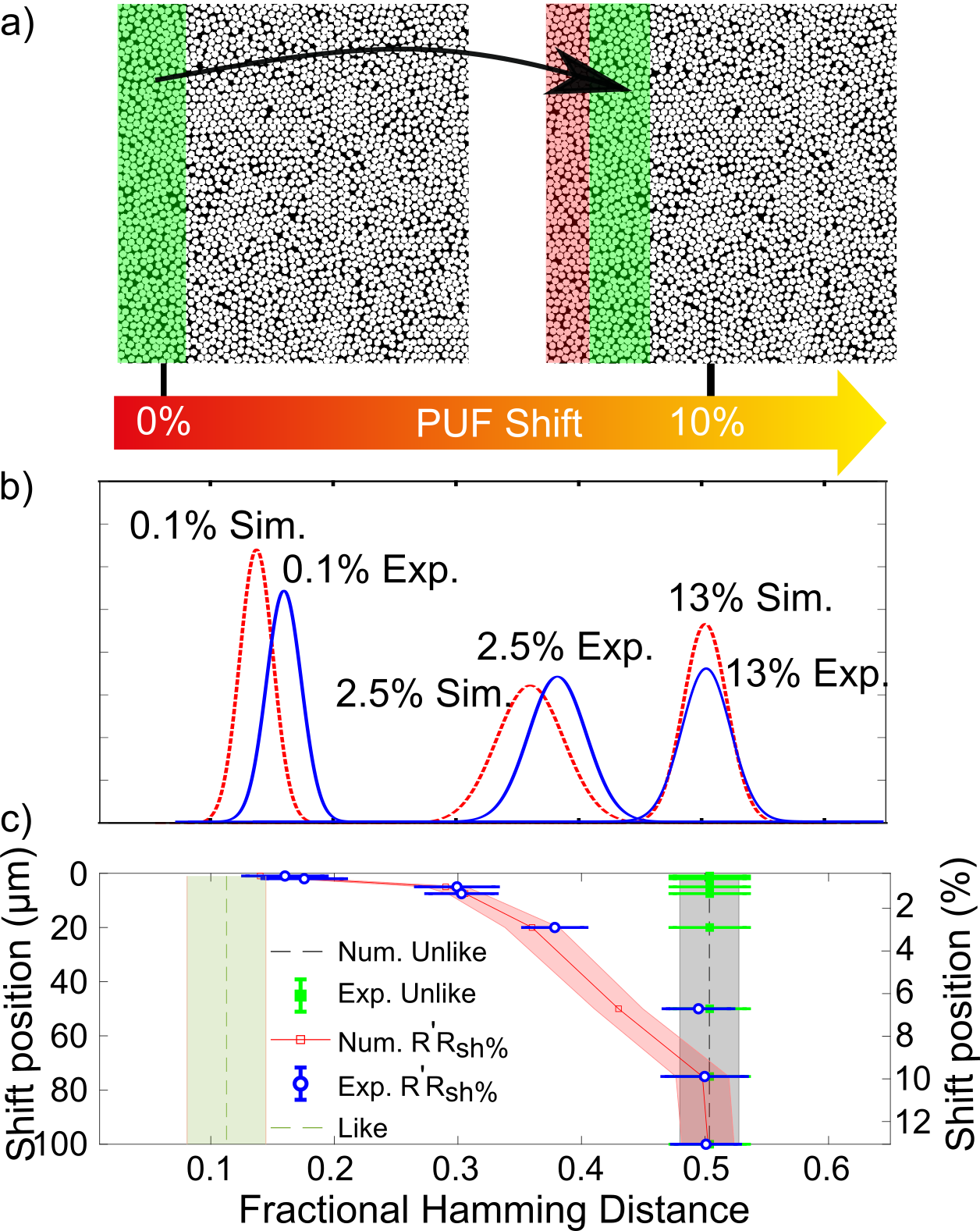}
\caption{a) Illustration of the optical PUF shift with respect to initial position and the probing/readout system. b) Numerical (red dashed lines) and experimental (blue solid lines) comparisons ($R'R_\text{sh\%}$) reported for a rigid shift along the $x$ direction, equal to \SI{0.1}{\percent}, \SI{2.5}{\percent}, and \SI{13}{\percent}. c) Trend for numerical and experimental ``unlike'' distributions (black ribbon and green dots respectively), the $R'R_\text{sh\%}$ comparison (red ribbon and blue hollow dots), and the experimental ``like'' (light green ribbon), for reference.}
\label{fig6}
\end{figure}
In this section, we study how the responses are affected by a misalignment of the illuminated region on the optical-PUF. This analysis has been carried out experimentally and verified by numerical calculation. The addressed case is of practical relevance assuming that the physical token must be manually inserted by a user in a slot for its optical readout. Numerically, we model the displacement by performing a translation of the  geometry along the $x$ axis, which is modeled with periodic boundary conditions for convenience (Figure \ref{fig6}a). Based on the so-called memory effect for speckle patterns \cite{feng1988correlations,yilmaz2021customizing}, a rigid shift of the PUF should correspond to a proportional shift of the response pattern if the displacement is small. In the following, the shift (sh\%) is evaluated as the ratio between the translation and the lateral size of the sample. We therefore expect that for small shifts, the $R'R_\text{sh\%}$ FHD distributions should remain almost unaffected as long as we re-align the optical responses using a registration algorithm (see Experimental Section). Numerical and experimental measurements are in good agreement and show that the compared distributions become independent (i.e., the responses are different and/or registration fails) after a shift of \SI{0.1}{\percent}, corresponding to \SI{\approx 1}{\micro\meter}, as reported in Figure \ref{fig6}b,c. 
\subsection{Sensitivity to Varying Scatterer Sizes}
\begin{figure}[!ht]
\centering
\includegraphics[width=\columnwidth]{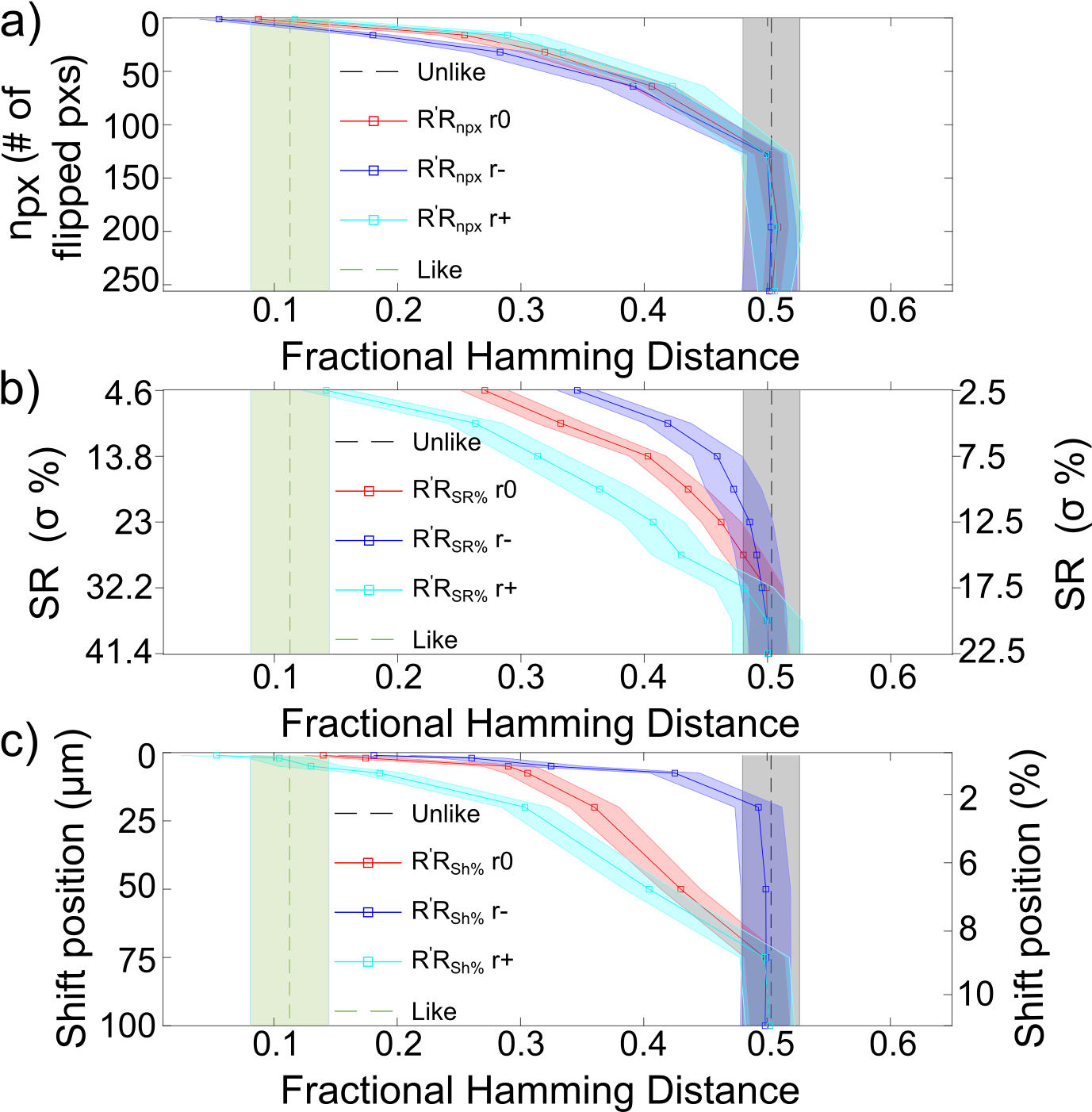}
\caption{The robustness/tolerance tunability based on the scatterers radius: a) increasing the scatterers radius the optical-PUF becomes more sensitive to a single flip in the $C_i$ challenge; (b-c) while decreasing it enhances the PUF sensitivity to scatterer misplacement or overall PUF misalignment (trends shown in the cyan, red, and blue ribbon lines, respectively). The black and the light green ribbons report the ``unlike'' and the experimental authentication threshold ``like'', for reference.}
\label{fig7}
\end{figure}
As a final test, we numerically study how the PUF sensitivity changes when the scatterer radius is either increased or decreased. Compared to the original hole radius of $r_0= \sim \SI{200}{\nano\meter}$, we test a reduced (subwavelength) value of $r_- = \SI{150}{\nano\meter}$, and $r_+ = \SI{400}{\nano\meter}$ (hole diameter > $\lambda$). The effect of this change is evaluated over the previous scenarios, including the challenge sensitivity, positional perturbation and rigid displacement tests. The results are summarized in Figure \ref{fig7}, showing that increasing the hole size leads to a slightly larger PUF sensitivity to small changes in the challenge (Figure \ref{fig7}a), but also to a larger tolerance to fabrication imperfections (Figure \ref{fig7}b), highlighting a trade-off between the strength and the unclonability for the considered 2D geometry. Similarly, the last panel shows how the hole radius affects also the overall misalignment tolerance of the PUF, which can be made significantly stricter by pushing the aperture size mode deeply into the sub-wavelength regime.
\section{Conclusions}

This paper quantified the sensitivity and unclonability of scattering-based optical physical unclonable functions in simulations and, for a few scenarios, also in experiments. We focused our study on 2D structures as they are very attractive for integrated devices in the visible and telecommunication range also for cryptographic applications.
We analyzed several relevant scenarios, including cloning attempts, challenge sensitivity, and the tolerance to device misalignment during optical readout.
Fabrication imperfections of few \si{\nano\meter}, or even the mismatch of one challenge pixel between the interrogation and the enrolled ones are sufficient to deny the authentication.
Evaluating the sensitivity of a PUF to small changes in the challenge pattern is also relevant to estimate the maximum CRP space that can be reasonably accessed using a PUF.
Finally, the misalignment test, relevant in a scenario where the PUF must be aligned into an optical readout device, shows that the tolerable misalignment is smaller than \SI{10}{\micro\meter}, for a realistic detector distance of \SI{5}{\milli\meter}.
These parameters, properly scaled for the PUF and readout system specs, should be taken into account when designing a readout device. 

At the same time, our results provide guidance on how to tune these values as needed towards either more robust or tolerant authentication devices, e.g., by acting on the radius of the holes in the structure.
Notably, our proposed numerical approach to test diffraction-based 2D optical PUFs is computationally efficient and flexible, allowing to investigate also other physical effects such as thermal expansion, mechanical stress, tampering attempts and readout noise or aberrations in future works.  

As a final remark, our results outline a general strategy to evaluate the sensitivity of optical physical unclonable functions under different scenarios, and provide a more quantitative ground to the general assumptions regarding their resilience against adversarial attacks.
Regardless of the specific 2D geometry considered here, these results are relevant also for more complex architectures with three-dimensional disorder, since the security level of 2D devices can be reasonably taken as a lower bound to the expected security of 3D PUFs.
Given the fast progress of advanced rigorous numerical methods, we envision that performing a similar analysis on a representative 3D arrangement of scatterers will soon be possible to test directly this assumption and quantify the security gain provided by multiple scattering also in other disordered 3D structures.
\section{Experimental Section}
\textbf{Sample fabrication}: The 2D optical-PUFs were experimentally fabricated by exploiting the irreproducible self-assembly of dielectric nanospheres (polystyrene) on glass substrates. The nanospheres were deposited via spin coating, in the regime of high spinning speeds (6000 rpm) where they form a monolayer with random arrangement (see Figure \ref{FigS3}a in SI). The initial diameter of the spheres (\SI{617}{\nano\meter}) was then reduced by Ar plasma to about \SI{400}{\nano\meter} to create a reflective mask (Figure \ref{FigS3}b) by evaporation of \SI{80}{\nano\meter} of Titanium (Figure \ref{FigS3}c). The spheres are then removed by sonication in isopropanol. This four-step process results in a 2D reflective perforated membrane with a random arrangement of nanoholes,  see Figure \ref{FigS3}d.\\
\textbf{Optical Setup}: A He-Ne laser beam ($\lambda = \SI{633}{\nano\meter}$, \SI{5}{\milli\watt}) propagates through lenses (L), polarizers (P), and a iris (I). The beam spot (magnified by a beam expander composed by a first lens L$_1$ the iris I and the magnification lens L$_2$) impinges on a digital micro-mirror device (DMD) used to generate the challenge ($C_i$) used to interrogate the scattering PUF-sample. The challenge $C_i$ is focused using an infinity-corrected lens (L$_{\infty}$ with focal length $f = \SI{200}{\milli\meter}$) into the objective back-focal plane (OBJ with $10\times$ magnification), which allows to de-magnify the challenges to a total area of $750 \times \SI{750}{\micro\meter\squared}$ corresponding to the physical extent of our experimental PUF. The optical pattern transmitted through the perforated membrane interferes in the far field to form a speckle pattern response ($R_i$), which is collected by lens in $2f$ configuration (L$_{2f}$) far away from the PUF. Finally, the speckle pattern is recorded by a $250 \times 250$ pixels camera (placed slightly off-axis to discard the ballistic signal) at 20 frames per second. A beam stabilizer is also included in the beam path, adjusting the beam position by means of a position detector (PD) controlling a piezoelectric mirror (PM), see Figure \ref{FigS1} in Supporting Information.\\
\textbf{Numerical simulations}: The Rayleigh-Sommerfeld (RS) method exploits a fast Fourier transformation operation evaluating the far field starting from the near field at a fixed distance along the light propagation direction $z$. Each random $C_i$ mask is created as a chessboard with size of $16 \times 16$ pixels filled with an equal number of ``on'' and ``off'' pixels placed randomly. The numerical sample contains a large number of scatterers ($\approx \num{2e6}$ for $f_\text{p} \approx 0.71$), and the generated responses are collected on a $250 \times 250$ pixel grid as in the experiments, at a distance of $z=\SI{5}{\milli\meter}$ and at an off-axis angle to avoid ballistic light.\\
\textbf{Image registration}: To properly evaluate the tolerance of the illumination and readout process to small shifts of the PUF, we apply an image registration algorithm to both experimental and numerical responses before calculating the FHD. Due to the high sensitivity of the Gabor hashing function to small changes in the responses, we find that registering speckle patterns has a relatively small impact when trying to recover a small misalignment, showing instead a much larger reduction of the FHD when applied to larger displacement values. Beyond a certain misalignment, however, the registration step itself will eventually fail, in which case we have left the alignment of $R'_i$ unmodified.
\section*{acknowledgements}
The authors thanks H. Cao for fruitful discussion and G. Roati and G. Del Pace for their help with the experimental equipment. This work was supported in part by the AFOSR/RTA2 (A.2.e.\ Information Assurance and Cybersecurity) project ``Highly Secure Nonlinear Optical PUFs''. GEL and FR thank the FASPEC (Fiber-Based Planar Antennas for Biosensing and Diagnostics) and the project "Complex Photonic Systems (DFM.AD005. 317). GEL also thanks the research project ``FSE-REACT EU'' financed by National Social Fund - National Operative Research Program and Innovation 2014-2020 (D.M. 1062/2021). Part of this work has been carried out at Nanofacility Piemonte INRiM, a laboratory supported by the ‘‘Compagnia di San Paolo’’ Foundation, and at the QR Laboratories, INRiM.
\section*{conflict of interest}
The authors declare no conflict of interest.
\section*{Data Availability Statement}
The data and the codes that support the findings of this study are available from the corresponding author upon reasonable request or at the following link over Zenodo repository 10.5281/zenodo.5986425. The \textit{``puffractio''} python code used to generate and process the numerical data is available at the following link \href{https://github.com/lpattelli/puffractio.git}{https://github.com/lpattelli/puffractio.git} on Github.
\section*{Supporting Information}
Supporting Information is available from the Wiley Online Library or
from the authors.
\printendnotes
%


%
\newpage

\setcounter{figure}{0}
\setcounter{equation}{0}
\renewcommand{\thefigure}{S\arabic{figure}}
\renewcommand{\theequation}{Ax\arabic{equation}}
\section*{\textbf{Supporting Information}}
\subsection*{Quantifying the Sensitivity and Unclonability of
Optical Physical Unclonable Functions}
\subsubsection*{Giuseppe Emanuele Lio$^{1,a), \dagger}$, Sara Nocentini$^{2,\dagger}$, Lorenzo Pattelli$^{2}$, Eleonora Cara$^{3}$, Diederik Sybolt Wiersma$^{4}$, Ulrich R\"{u}hrmair$^{5,b)}$, Francesco Riboli$^{6,c)}$.}
\begin{figure}[!ht]
\includegraphics[width=\columnwidth]{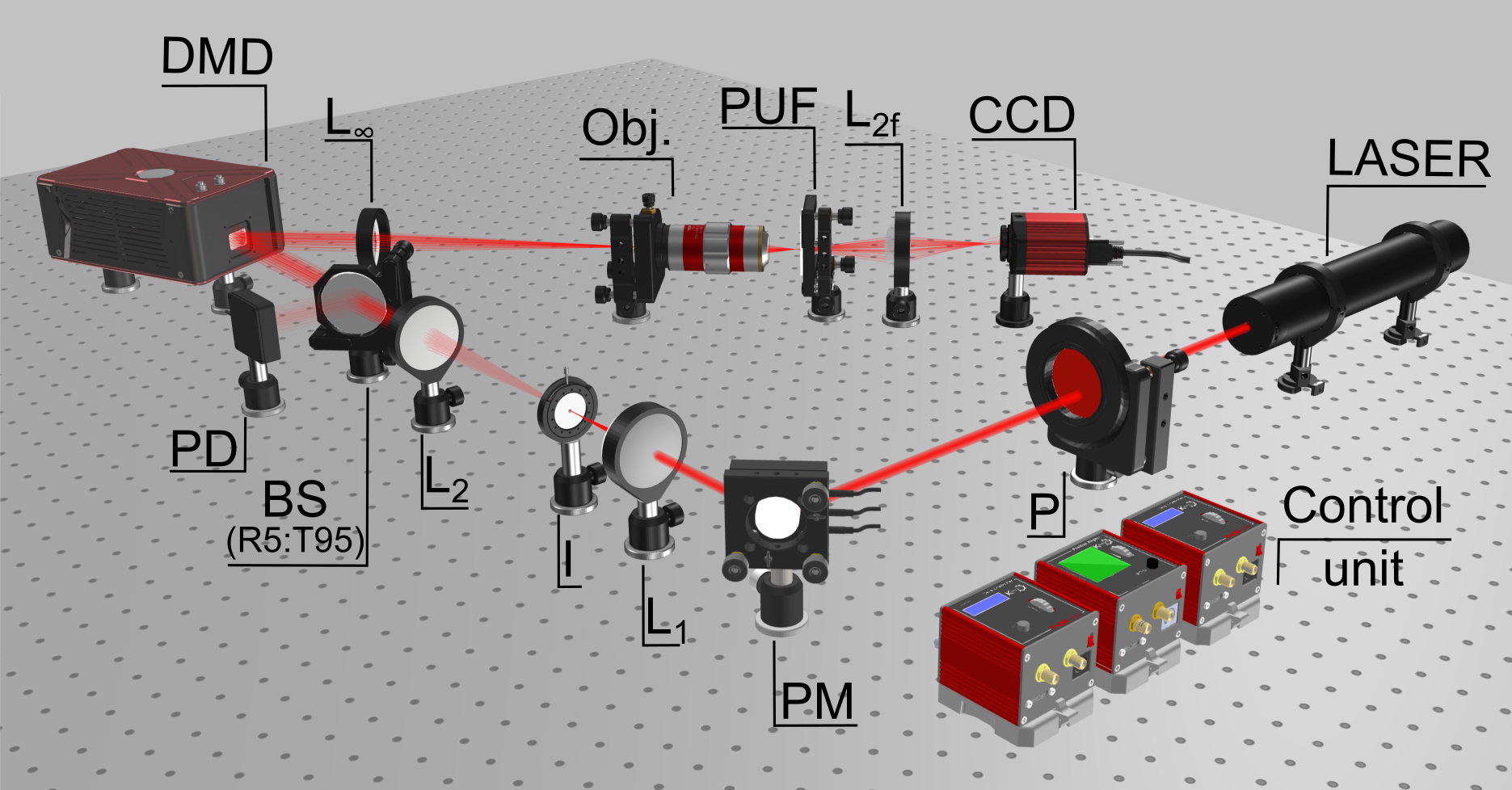}
\caption{Representative sketch of the experimental setup used to probe the PUF with the challenges $C_1, \dots, C_i$ and to collect the responses $R_1, \dots, R_i$.} 
\label{FigS1}
\end{figure}  
\subsection*{Challenges pixel size}
\begin{figure}[!ht]
\includegraphics[width=\columnwidth]{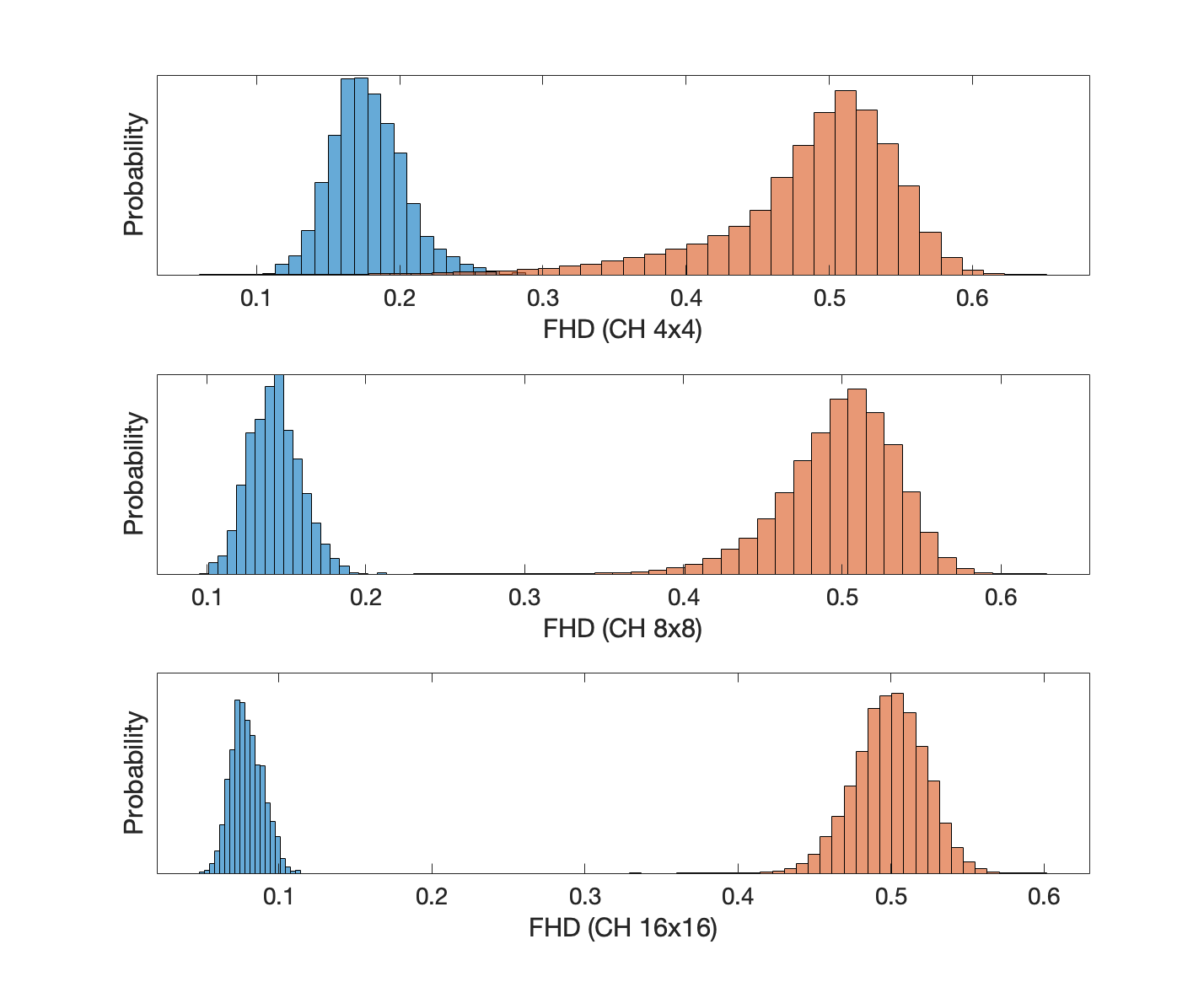}
\caption{FHD ``like'' and ``unlike'' distributions referred to a set of 2000 challenges with $4\times4$, $8\times8$, and $16\times16$ macro-pixels.} 
\label{FigS2}
\end{figure}

\begin{figure}[!ht]
\includegraphics[width=\columnwidth]{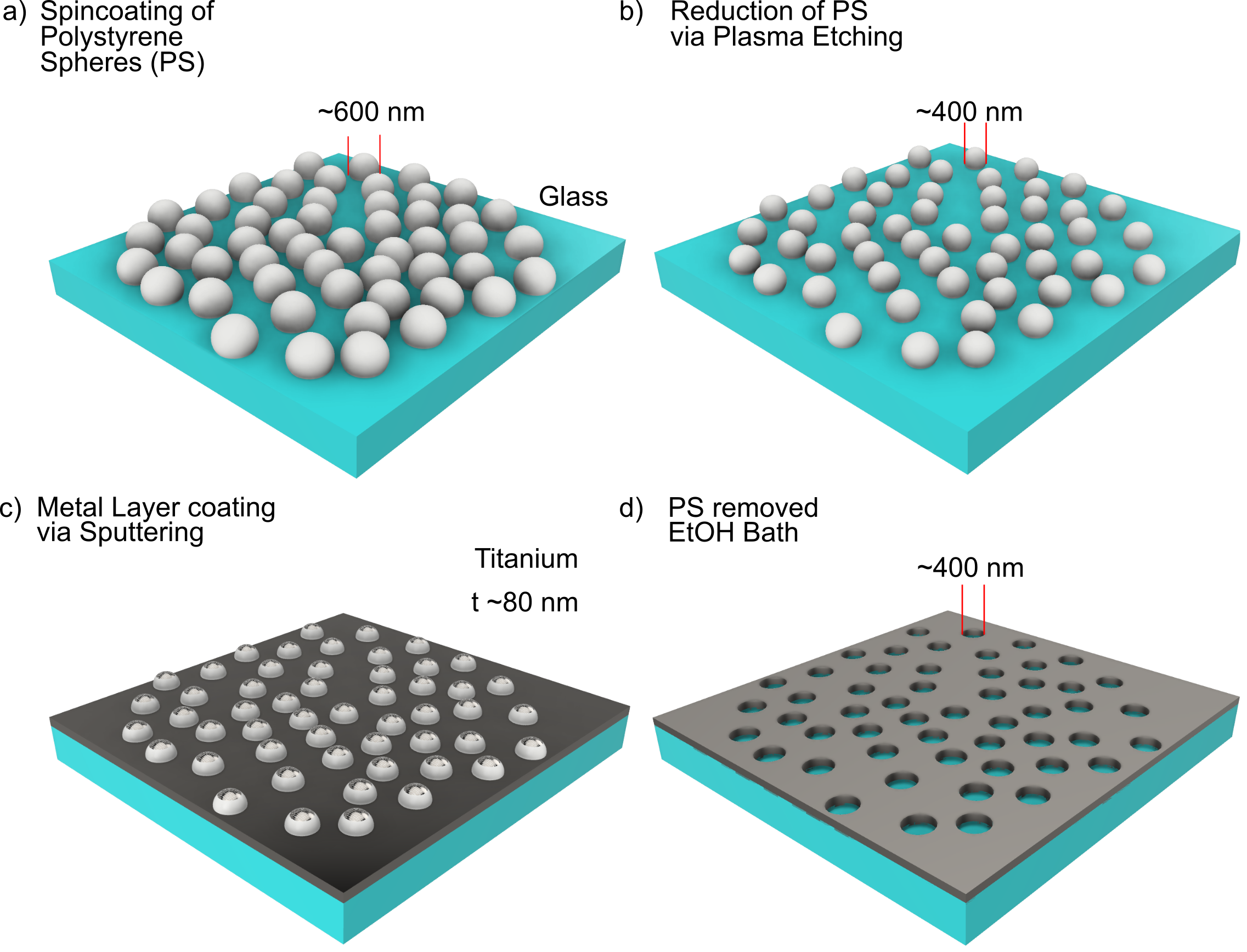}
\caption{Optical-PUF fabrication process: a) Spin coating of polystyrene spheres (PS), $D\approx \SI{600}{\nano\meter}$ on microscope glass. b) Plasma etching of the PS and diameter reduction ($D \approx \SI{400}{\nano\meter}$). c) Titanium thin layer sputtered to cover the substrate ($t\approx$ \SI{80}{\nano \meter}). d) PS removal by ethanol bath.} 
\label{FigS3}
\end{figure}

The number of macro pixels for the challenges has been selected experimentally using \num{2000} challenges on the real optical-PUF with a different macro-pixels arrangement. To this end, challenges with $4\times4$, $8\times8$, and $16\times16$ macro-pixels have been used. As reported in Figure \ref{FigS2} using too few macro-pixels results in large and broad ``unlike'' and ``like'' distributions. The intermediate case shows a slightly better distribution of ``unlike'' and ``like'' distributions, while using $16\times16$ challenge patterns the produced keys shown narrowed ``unlike'' and ``like'' FHD distributions with a high stability, well separated and a ``like'' FHD distribution close to the possible threshold ($ \expval{p}\approx0.1$). Furthermore, the challenges with $16\times16$ macro-pixels represent a suitable configuration to have a response sensitivity to a single bit flip modifying just an area equal to $1/256$ of the illumination pattern instead of the other cases, namely $4\times4$ and $8\times8$,  where a single flip corresponds to modify $1/16$ or $1/64$ of the challenge area respectively. On the other hand, as shown in the subsection 3.1, using challenge patterns composed of $32\times32$ or more macro-pixels does not to satisfy the condition that a single bit flip in the challenge should correspond to a shift of the FHD distribution $R'R_{1px}$ large enough to avoid overlaps with the ``like'' standard deviation.
\newpage
\subsection*{Sample fabrication}

%
\end{document}